\documentclass[sigconf]{acmart}
\usepackage{algorithm}
\usepackage{array}  
\usepackage{algorithmic}
\usepackage{booktabs}
\usepackage{multirow}
\usepackage[most]{tcolorbox}
\usepackage{enumitem}  
\AtBeginDocument{%
  }

\setcopyright{acmlicensed}
\copyrightyear{2025}
\acmYear{2025}
\acmDOI{XXXXXXX.XXXXXXX}
\acmConference[Conference acronym 'XX]{Make sure to enter the correct
  conference title from your rights confirmation email}{June 03--05,
  2018}{Woodstock, NY}
\acmISBN{978-1-4503-XXXX-X/2018/06}
\newcommand{\methodname}{Agent4SR}

\begin{document}
\begin{sloppypar}

\title{LLM-Based User Simulation for Low-Knowledge Shilling Attacks on Recommender Systems}
\author{Shengkang Gu}
\authornote{Equal contribution.}
\author{Jiahao Liu}
\authornotemark[1]
\affiliation{
  \institution{Fudan University}
  \city{Shanghai}
  \country{China}
}
\email{gusk24@m.fudan.edu.cn}
\email{jiahaoliu23@m.fudan.edu.cn}


\author{Dongsheng Li}
\affiliation{
  \institution{Microsoft Research Asia}
  \city{Shanghai}
  \country{China}
}
\email{dongsli@microsoft.com}

\author{Guangping Zhang}
\affiliation{
  \institution{Fudan University}
  \city{Shanghai}
  \country{China}
}
\email{gpzhang20@fudan.edu.cn}

\author{Mingzhe Han}
\affiliation{
  \institution{Fudan University}
  \city{Shanghai}
  \country{China}
}
\email{mzhan22@m.fudan.edu.cn}

\author{Hansu Gu}
\affiliation{
  \institution{Independent}
  \city{Seattle}
  \country{United States}
}
\email{hansug@acm.org}

\author{Peng Zhang}
\affiliation{
  \institution{Fudan University}
  \city{Shanghai}
  \country{China}
}
\email{zhangpeng_@fudan.edu.cn}

\author{Ning Gu}
\affiliation{
  \institution{Fudan University}
  \city{Shanghai}
  \country{China}
}
\email{ninggu@fudan.edu.cn}

\author{Li Shang}
\affiliation{
  \institution{Fudan University}
  \city{Shanghai}
  \country{China}
}
\email{lishang@fudan.edu.cn}

\author{Tun Lu}
\affiliation{
  \institution{Fudan University}
  \city{Shanghai}
  \country{China}
}
\email{lutun@fudan.edu.cn}


\renewcommand{\shortauthors}{Shengkang Gu et al.}

\begin{abstract}
Recommender systems (RS) are increasingly vulnerable to shilling attacks, where adversaries inject fake user profiles to manipulate system outputs. Traditional attack strategies often rely on simplistic heuristics, require access to internal RS data, and overlook the manipulation potential of textual reviews. In this work, we introduce Agent4SR, a novel framework that leverages Large Language Model (LLM)-based agents to perform low-knowledge, high-impact shilling attacks through both rating and review generation. Agent4SR simulates realistic user behavior by orchestrating adversarial interactions—selecting items, assigning ratings, and crafting reviews—while maintaining behavioral plausibility. Our design includes targeted profile construction, hybrid memory retrieval, and a review attack strategy that propagates target item features across unrelated reviews to amplify manipulation. Extensive experiments on multiple datasets and RS architectures demonstrate that Agent4SR outperforms existing low-knowledge baselines in both effectiveness and stealth. Our findings reveal a new class of emergent threats posed by LLM-driven agents, underscoring the urgent need for enhanced defenses in modern recommender systems.
\end{abstract}


  

\received{20 February 2007}
\received[revised]{12 March 2009}
\received[accepted]{5 June 2009}


\maketitle

\section{Introduction}

With the rapid growth of e-commerce, \textbf{R}ecommender \textbf{S}ystems (RS) have become essential for personalized content delivery~\cite{ricci2010introduction,liu2023personalized,liu2022parameter,liu2023autoseqrec,liu2023recommendation,liu2023triple}. For users, RS leverage interaction data, particularly ratings and reviews, to predict their preferences for specific items~\cite{liu2025filtering,liu2025mitigating,liu2025enhancing1,han2025fedcia}. For content providers (\textit{e.g.}, online retailers), RS help match their items with interested users, thereby enhancing item exposure and driving revenue growth~\cite{boutilier2023modelingrecommenderecosystemsresearch, kiyohara2025policydesigntwosidedplatforms}.
As a result, malicious actors may attempt to manipulate RS to maximize their own interests. A common tactic is \textit{\textbf{shilling attack}}, where malicious user profiles with fabricated interaction data are injected into the system’s training data to influence the outcome of RS~\cite{Sundar2020UnderstandingShillingAttacksandTheirDetectionTraits:AComprehensiveSurvey, lam2004Shillingrecommendersystemsforfunandprofit, Lin_2020, Lin_2024}.

\begin{figure}[t]
  \centering
  \includegraphics[width=\linewidth]{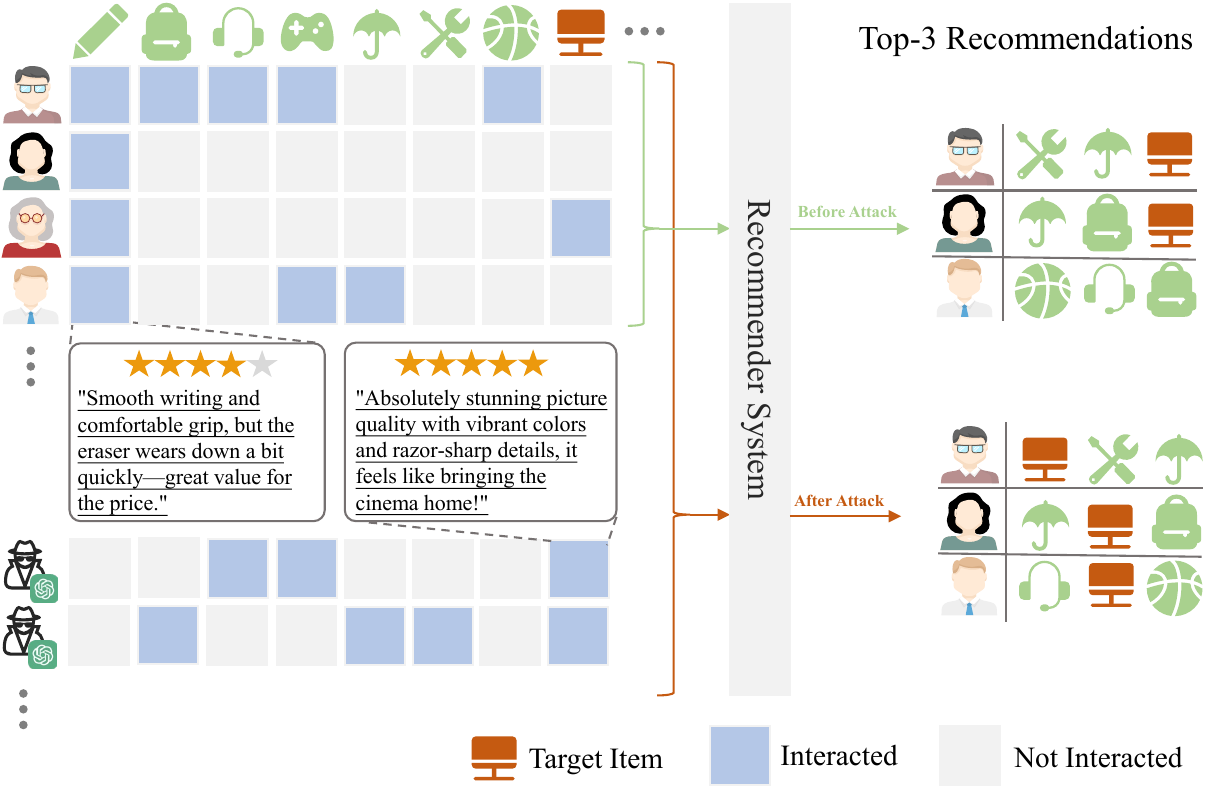}
  \caption{Overview of \methodname{}. As shown, injecting fake user profiles generated by LLM-based user agents manipulates recommendations and substantially boosts the target item’s ranking.}
  \label{fig:intro}
  \Description{Introduction to shilling attack.}
\end{figure}

The essence of shilling attacks lies in exploiting distributional differences between fake and genuine users to influence RS performance. \textbf{Existing shilling attack methods exhibit several limitations:}
(1) Conventional shilling attacks assign extreme ratings to target items: maximal for promotion (push attack) or minimal for demotion (nuke attack). For non-target items, attackers typically select them randomly and assign ratings using strategies like random values or item averages to mimic genuine user behavior~\cite{lam2004Shillingrecommendersystemsforfunandprofit, omahony2005bandwagon, burke2005limited, mobasher2007attacks, williams2006detection, Hurley2009statistical}. These rule-based profiles often lack diversity, making them easily detected as anomalies. 
(2) Some methods enhance camouflage by integrating additional external knowledge. For example, methods based on \textbf{G}enerative \textbf{A}dversarial \textbf{N}etworks (GAN)~\cite{goodfellow2014generativeadversarialnetworks} generate fake user profiles by sampling patterns from genuine user interaction data~\cite{wilson2013power, seminario2016nuking, tang2020revisiting, song2020PoisonRec, Christakopoulou2018AdversarialRA, Christakopoulou2019adversarial, Lin_2020, Lin_2024}. These models produce stealthier fake profiles and train a discriminator on both real and synthetic interaction data. Their main limitation lies in their heavy reliance on internal RS data (\textit{e.g.}, genuine users' interactions), which is typically inaccessible in real-world attack scenarios.
(3) Existing shilling attacks primarily manipulate rating data while often neglecting user reviews, which can carry greater risk due to their nuanced and semantically rich content~\cite{yang2023incorporated, chiang2023shilling}. With advances in text mining and the increasing availability of user reviews, leveraging textual information in RS has become both common and effective~\cite{gao2020Set-sequence-graph:Amulti-viewapproachtowardsexploitingreviewsforrecommendation}. Therefore, it is essential to investigate review-based attack strategies that complement rating-based approaches and to assess their potential impact on RS.

In recent years, \textbf{L}arge \textbf{L}anguage \textbf{M}odels (LLMs) have demonstrated strong capabilities in human-level reasoning, decision-making, and natural language generation~\cite{lee2024Reasoningabilitiesoflargelanguagemodels:In-depthanalysisontheabstractionandreasoningcorpus}. This progress has motivated the development of LLM-based agents that simulate user behaviors using only minimal system-specific data. Consequently, several studies have proposed LLM-based agents to simulate users in RS~\cite{recagent, zhang2024Ongenerativeagentsinrecommendation, zhang2024agentcf, liu2025enhancing}. These agents serve multiple purposes, including bridging the gap between offline evaluation and online performance and modeling complex user behavior~\cite{zhang2024Ongenerativeagentsinrecommendation, huang2024recommenderaiagentintegrating}. However, the openness of RS exposes them to \textbf{malicious user agents}, either externally injected or derived from internally compromised LLM-based agents, both of which can severely undermine system integrity. Synthetic agents embedded with targeted malicious content can induce biased behaviors, including preferential promotion of specific vendors or suppression of certain product categories. By participating in activities such as rating and reviewing, these agents can manipulate recommendation outcomes in ways that serve the interests of their orchestrators.

In this paper, we propose \textbf{\methodname{}}, a novel low-knowledge shilling attack framework that employs LLM-based agents to manipulate RS. We inject a set of fake user agents into RS, which engage in adversarial yet human-like interactions by selecting items, rating them, and posting reviews to generate synthetic data that blends with real user interactions. This manipulation distorts the recommendation outcomes. For instance, in a promotion attack, the target item receives increased exposure, as illustrated in Figure~\ref{fig:intro}.

Specifically, we designed tailored attack strategies for each component of the fake user agent.
We developed a method for the profile module that infers and validates agent personality traits based on the target item and incorporates mechanisms to enhance diversity.
In parallel, the memory module employs retrieval strategies that consider both relevance and recency.
In the action module, beyond generating ratings, the agent is also capable of producing reviews, for which we designed a target feature propagation strategy.
We simulated fake user interactions in the RS by structuring the information flow from the system into the agent.

We validated the effectiveness of \methodname{} across different types of RS, including those using ratings alone and those augmented with reviews. \methodname{} consistently outperformed existing low-knowledge attacks, highlighting its strong attack capability. The fake user agent exhibited strong behavioral resemblance to genuine users in our experiments, making it difficult to detect even with advanced detection algorithms. We conducted ablation experiments to evaluate the contribution of each component in \methodname{}. We also performed additional studies to investigate how fake users of varying scales, targeting items with different popularity levels, affect users across different activity tiers.

To the best of our knowledge, \methodname{} is the first to leverage LLM-based agents to simulate users for attacking RS, and we demonstrate that the resulting attacks exhibit both substantial harm and stealth. By leveraging the \textbf{human-like reasoning}, \textbf{rich world knowledge}, and \textbf{natural language generation capabilities} of LLMs, \methodname{} overcomes key limitations of existing approaches. \methodname{} introduces a paradigm shift in shilling attacks, moving from rule-based heuristics to cognitively simulated agent behavior. As the use of LLM-based agents becomes more widespread, our objective is to highlight emerging security risks posed by such agents acting like genuine users.


\section{Related Work}

\subsection{Shilling Attacks against RS}
Numerous methods have been proposed for black-box shilling attacks against RS. Early studies explored strategies that utilize global statistical information to generate fake user profiles for push attacks (\textit{e.g.}, random, bandwagon, segmented) and nuke attacks (\textit{e.g.}, reverse bandwagon, love/hate)~\cite{lam2004Shillingrecommendersystemsforfunandprofit, omahony2005bandwagon, burke2005limited, mobasher2007attacks, Sundar2020UnderstandingShillingAttacksandTheirDetectionTraits:AComprehensiveSurvey}.
To improve attack stealth, heuristic techniques such as noise injection, user shifting, target shifting, and popularity-based averaging have been proposed to obscure aggressive rating patterns and evade detection~\cite{williams2006detection, Hurley2009statistical}. These methods represent simple low-knowledge black-box attacks that rely on predefined rules and \textbf{fail to capture the behavioral complexity of genuine users in personalized contexts} which limits their stealthiness.

Several studies assume that attackers have access to more knowledge, which can be used to diversify malicious user data and enhance its stealth against detection. For example, some attacks leverage \textbf{detailed interaction features} such as item rating counts and user or item in-degree centrality to identify and target influential nodes in the system~\cite{seminario2016nuking, wilson2013power}. Other work considers scenarios where \textbf{partial knowledge of the RS structure is available} and uses surrogate models to predict attack outcomes and guide item selection~\cite{tang2020revisiting}. Reinforcement learning has also been used to train an agent that inject fake interactions and optimize attack strategies based on \textbf{periodic feedback from the RS}, although such feedback is rarely available in practice~\cite{song2020PoisonRec, chiang2023shilling}.
The emergence of GANs has further advanced this line of work by enabling the automatic generation of fake user profiles. One approach applies DCGAN to generate initial profiles, trains a discriminator \textbf{to distinguish real from fake ratings}, and refines the profiles using zero-order optimization~\cite{Christakopoulou2018AdversarialRA, Christakopoulou2019adversarial, radford2016unsupervisedrepresentationlearningdeep}. Another line of work improves GAN-based attacks by sampling genuine user templates from \textbf{the rating matrix} and using a generator to produce fake ratings, with a discriminator optimizing a global objective that integrates multiple loss functions~\cite{Lin_2020, Lin_2024}.
R-Trojan~\cite{yang2023incorporated} generates fake profiles by combining user ratings with fine-tuned review generation. However, its review generation \textbf{lacks explicit attack objectives} and still relies on internal data such as \textbf{the rating matrix} for training, limiting its applicability.

\subsection{LLM-Based User Agents in RS}
LLM-based agents in RS can be broadly categorized into two types. The first leverages LLMs to generate or enhance recommendations, referred to as recommendation agents~\cite{huang2024recommenderaiagentintegrating, Shi2024Large, wang2024recmindlargelanguagemodel, Wang_2024, zhang2024prospectpersonalizedrecommendationlarge, zhang2024recommendation, Zhao_2024}. The second focuses on simulating user behavior through LLMs, known as user agents. While some studies model user dialogues in conversational recommendation settings~\cite{friedman2023leveraginglargelanguagemodels, kim2024stopplayingguessinggame, Wang_2023, zhu2024howreliable, zhu2024llmbasedcontrollablescalablehumaninvolved}, our work emphasizes modeling user actions such as rating and reviewing to reflect realistic interaction behavior in RS.

RecAgent~\cite{recagent} and Agent4Rec~\cite{zhang2024Ongenerativeagentsinrecommendation} employ LLM-based agents that integrate \textbf{profile}, \textbf{memory}, and \textbf{action} modules to simulate user decision-making processes in RS. RAH~\cite{shu2024rah} introduces LLM-based multi-agents that serve both as recommendation generators and as simulators of user behavior, bridging users and RS. FLOW~\cite{cai2025agenticfeedbackloopmodeling} facilitates collaboration between recommendation and user agents via a feedback loop. Zhang et al.~\cite{zhang2024llmpoweredusersimulatorrecommender} integrate explicit preferences, LLM-based sentiment analysis, human-in-the-loop modeling, and statistical frameworks to simulate user interactions more robustly. AgentCF~\cite{zhang2024agentcf} introduces a co-learning framework that models users and items as agents and optimizes them jointly. AgentCF++~\cite{liu2025enhancing} extends this framework to support cross-domain transfer and incorporate popularity-aware modeling. Together, these works highlight the potential of LLMs for simulating human-like user behavior in RS.

\section{Preminilaries}
In this section, we formally define the shilling attack problem and introduce key related concepts.

\subsection{Definition of Shilling Attack}
Shilling attacks involve injecting fake user profiles into RS to manipulate recommendation outcomes.

Let the input to the RS be a user-item interaction dataset $\mathcal{D} = \{D_{u,i} \mid u \in U, i\in I\}$, where $U = \{u_1, \dots, u_m\}$ is the set of genuine users, and $I = \{i_1, \dots, i_n\}$ is the set of items. Each $D_{u,i} \in \mathcal{V}$ denotes the interaction between user $u$ and item $i$, where $\mathcal{V}$ is the domain of interaction values (\textit{e.g.}, numerical ratings $\mathbb{R}$ or textual reviews $\mathcal{T}$).

An attacker introduces a set of fake users $U_a$, and constructs corresponding fake interaction data $\mathcal{D}_a = \{ D_{u_a,i} \mid u_a \in U_a, i \in I\}$. The manipulated input to the system becomes $\mathcal{D'} = \mathcal{D} \cup \mathcal{D}_a$.

The attacker aims to manipulate the recommendation algorithm $\Phi: \mathcal{D} \to \mathcal{O}$, where $\mathcal{O}$ denotes the system output (\textit{e.g.}, predicted rating matrix or top-N list), to significantly alter the recommendation performance of the target item $i_t \in I$.

In the case of a push attack, the attacker aims to improve the recommendation outcome for the target item, defined as:
\begin{equation}
    \Delta \Phi(i_t) = \Phi(\mathcal{D'})_{i_t} - \Phi(\mathcal{D})_{i_t}
\end{equation}
Here, $\Phi(\mathcal{D}){i_t}$ and $\Phi(\mathcal{D'}){i_t}$ denote the recommendation outcome for item $i_t$ before and after the attack, respectively (\textit{e.g.}, predicted score or ranking position).

The shilling attack can be formulated as the following optimization problem:
\begin{equation}
    \mathop{\text{\Large max}}\limits_{U_a,\, \mathcal{D}_a} \Delta \Phi(i_t)
\end{equation}

\subsection{Anatomy of Shilling Attacks}

The interacted items of each fake user $u_a \in U_a$ are typically partitioned into three categories, each governed by a different distribution determined by the attack method.

\paragraph{Target Items.} For the target item $i_t$, fake user interactions are generated to follow a biased distribution $P_{target}$ reflecting the attacker’s adversarial intent:
\begin{equation}
    D_{U_a,i_t} \sim P_{target}
\end{equation}
For example, in a push attack, fake users may assign high ratings and generate positive reviews for the target item. In a nuke attack, they may assign low ratings and post negative reviews.

\paragraph{Selected Items.} Beyond the target item, the attacker typically selects a set of items $I_s \subseteq I \setminus \{i_t\}$ to emulate genuine user behavior. For each selected item $i_s \in I_s$, fake user interactions are generated to follow a distribution:
\begin{equation}
    D_{U_a,i_s} \sim P_{selected}
\end{equation}
Here, $P_{selected}$ is designed to approximate typical user behavior and improve the stealth of fake profiles. For instance, in a bandwagon attack, fake users may assign high ratings to popular items to resemble normal user patterns.

\paragraph{Filler Items.} The filler item set is defined as $I_f \subseteq I \setminus (\{i_t\} \cup I_s)$. For each filler item $i_f \in I_f$, the fake user's interaction is generated to follow:
\begin{equation}
    D_{U_a,i_f} \sim P_{filler}
\end{equation}
where $P_{filler}$ is defined according to the specific attack method. The filler set may vary across fake users to increase interaction diversity and reduce detectability.

The remaining items, with which the fake user does not interact, are denoted as $I_{\phi} = I \setminus ({i_t} \cup I_s \cup I_f)$.

In essence, different shilling attack methods differ in how they construct $U_a$, choose $I_s$ and $I_f$, and define the corresponding interactions $D_{U_a,i_t}$, $D_{U_a,i_s}$, and $D_{U_a,i_f}$. These components are collectively optimized to maximize $\Delta \Phi(i_t)$. Equally important is ensuring that fake user behavior is well-camouflaged within the distribution of genuine users, thereby enhancing their resistance to detection.

\section{Methodology}
\begin{figure*}[h]
  \centering
  \includegraphics[width=\linewidth]{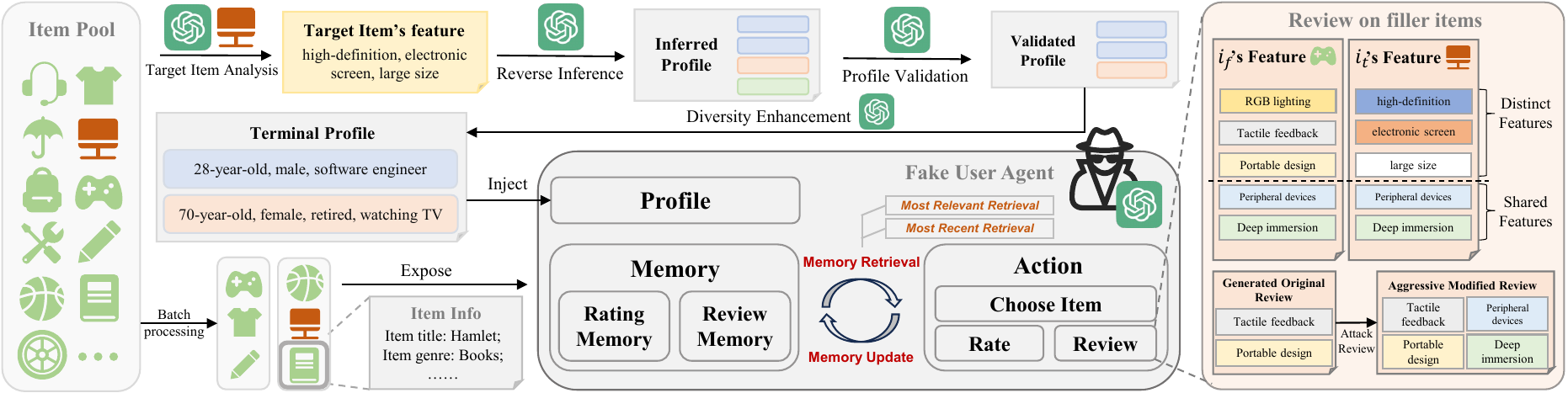}
  \caption{The overall framework of \methodname{}. It starts with target item analysis and profile inference, followed by profile validation and diversification. Fake user agents interact with filler items using hybrid memory retrieval, while a target feature propagation strategy embeds key target attributes into reviews on filler items.}
  \label{fig:method}
  \Description{Method with LLM-based Agent shilling attack.}
\end{figure*}

This section details the design of \methodname{}, which employs an LLM-based user agent with a modular architecture comprising a \textbf{profile}, \textbf{memory}, and \textbf{action} module~\cite{zhang2024Ongenerativeagentsinrecommendation, recagent}. These components are integrated to support targeted and effective shilling attacks. The method is structured around three core design components:

\begin{enumerate}[left=0.5em]
    \item Adversarial profile construction via target-guided inference, followed by validation and diversity enhancement.
    \item Hybrid memory retrieval based on relevance and recency.
    \item Multifaceted interaction attack enhanced by target feature propagation strategy.

\end{enumerate}
The overall framework is illustrated in Figure~\ref{fig:method}.

\subsection{Modular Structure of Fake User Agent}
We formally represent the fake user agent as a triplet $\text{Agent} = (\mathcal{P}rofile, \mathcal{M}emory, \mathcal{A}ction)$, comprising three core modules:
\begin{itemize}[left=0.5em]
    \item \textbf{Profile Module}: Stores personalized attributes that define the agent’s identity and preferences, such as age, gender, occupation, user-specific interests, and interaction habits (\textit{e.g.}, rating or reviewing frequency).
    \item \textbf{Memory Module}: Stores the agent’s historical interactions, including ratings and reviews. It employs two retrieval methods: a relevance-based retriever and a recency-based retriever, to distill key information and support decision making.
    \item \textbf{Action Module}: Defines the agent's behavior, including item selection, rating generation, and review generation.
\end{itemize}

\subsection{Target-Aware Profile Construction}
To construct a set of fake user agents whose preferences align with a given target item, we propose a target-driven method that reversely infers user profiles from the target item’s characteristics, followed by validation and diversification.

\subsubsection{Target Item Feature Analysis}
Given a target item $i_t$, we construct a prompt containing its basic metadata (\textit{e.g.}, title and category), and invoke the LLM to generate an analysis of its characteristics $\mathcal{A}_t = \text{Prompt}{\text{\tiny LLM}}(i_t)$ based on its pretrained world knowledge.

\subsubsection{Profile Inference from Target Item}
Based on the item analysis $\mathcal{A}_t$ and the attack type $d \in \{+,-\}$ (push or nuke), we construct the prompt for profile inference.
For example, in a push attack on the item titled "\textit{High-Definition Large-Size Electronic Display Screen}", the prompt is constructed as:

\begin{tcolorbox}[colback=gray!5, colframe=black!40, title=Example Prompt for Profile Inference]

\textbf{You currently have a strong preference for} High-Definition Large-Size Electronic Display Screen.

\textbf{The information about this item is as follows:} It features an ultra-large 32-inch curved design with high resolution and exceptional clarity. It is suitable for gaming, multitasking, office work, and watching TV or other media.

\textbf{Based on this item description, please infer your potential personality profile, including:}
\begin{itemize}[left=0.5em]
  \item \textbf{Gender, age, and occupation}
  \item \textbf{Personal interests or dislikes}
  \item \textbf{Interaction habits} (\textit{e.g.}, \textbf{how frequently you provide reviews or ratings})
\end{itemize}
\end{tcolorbox}

The LLM output from this prompt defines the Profile module $\mathcal{P}^{(i_t)}$, comprising user interests, interaction habits, and other personality traits.
\begin{equation}
\label{gen_profile}
    \mathcal{P}^{(i_t)} = Prompt_{\text{\tiny LLM}}({\mathcal{A}_t, d})
\end{equation}

\subsubsection{Profile Consistency Validation}
To verify that the generated profile $\mathcal{P}^{(i_t)}$ logically aligns with the target item $i_t$, we introduce a validation step inspired by the encoding-decoding architecture in Seq2Seq models~\cite{Sutskever2014Sequencetosequencelearningwithneuralnetworks}. The profile generation serves as the encoding phase, while validation corresponds to decoding. Specifically, we issue a secondary prompt asking the fake user agent, based on its current profile, whether it would genuinely prefer the target item, as in the case of a push attack. A negative response (\textit{i.e.}, $\text{Validate}(\mathcal{P}^{(i_t)})=0$) triggers profile regeneration to ensure consistency with the attack objective.
\begin{equation}
\label{valid_profile}
\text{Validate}(\mathcal{P}^{(i_t)}) = 
\begin{cases}
1, & \text{if } Prompt_{\text{\tiny LLM}}(\mathcal{P}^{(i_t)}, i_t, d){=}\text{Yes} \\
0, & \text{otherwise}
\end{cases}
\end{equation}

\subsubsection{Profile Diversity Enhancement}
To enhance diversity among fake user agents and reduce the risk of detection caused by overly concentrated interaction patterns, we leverage the LLM to select a subset of profiles that maximizes diversity from the candidate pool:
\begin{equation}
\label{diver_profile}
    \{\mathcal{P}^{(i_t)}_k\}_{k=1}^{N} = Prompt_{\text{\tiny LLM}}(\{\mathcal{P}^{(i_t)}_k\}_{k=1}^{M})
\end{equation}
where $M$ is the number of candidate profiles generated, and $N$ is the number of fake user agents to be injected.

Overall, in a push attack, each fake user agent is individually crafted to resemble a genuine user who appears to hold authentic preferences for the target item. At the population level, the fake users exhibit diversity comparable to real-world user distributions. When interacting with the RS, their behaviors appear highly plausible and are difficult to detect.

\subsection{Memory Management and Hybrid Retrieval}
The Memory module stores the fake user agent’s historical interactions, consisting of rating records $\mathcal{M}_{R} = {(i_m, \mathbb{R}_m, t_m)}$ and review records $\mathcal{M}_{T} = {(i_n, \mathcal{T}_n, t_n)}$, where $i$ denotes the item, $\mathbb{R}$ the rating, $\mathcal{T}$ the review, and $t$ the timestamp. The module supports memory updates and retrieval to guide downstream decisions.

\subsubsection{Memory Update Mechanism}
Upon executing an action, the agent updates its memory by recording the new rating $(i_q, \mathbb{R}_{new}, t_{now})$ in $\mathcal{M}_{R}$ and the corresponding review $(i_q, \mathcal{T}_{new}, t_{now})$ in $\mathcal{M}_{T}$. To avoid excessive memory accumulation, only a fixed number of recent entries may be retained.

\subsubsection{Memory Retrieval Mechanism}
Given a query item $i_q$, we perform two types of retrieval: relevance-based and recency-based.

(1) \textbf{Relevance-based retrieval} selects the top-$K$ semantically relevant entries from the full memory set $\mathcal{M} = \mathcal{M}_{R} \cup \mathcal{M}_{T}$:
\begin{equation}
\label{relevancebasedretrieval}
    \mathcal{M}_{rel} = \{m_j \in \mathcal{M} \mid sim(i_q,i_j) \geq Top_K(\{sim(v_q,v_k)\}_{k=1}^{\lvert \mathcal{M} \rvert})\}
\end{equation}
where $\text{sim}(\cdot)$ denotes semantic similarity based on item information. Specifically, item semantics are extracted using a Transformer-based language model, which encodes textual metadata (\textit{e.g.}, item titles) into dense vector representations. Cosine similarity is then computed between these embeddings.

(2) \textbf{Recency-based retrieval} selects the $M$ most recent interaction records:
\begin{equation}
\label{recencybasedretrieval}
    \mathcal{M}_{rec} = \{ m_i \in \mathcal{M} \mid t_i \geq Top_M(\{t_j\}_{j=1}^{\lvert \mathcal{M} \rvert})\}
\end{equation}

In summary, the final retrieved memory set $\mathcal{M}_{ret} = \mathcal{M}_{rel} \cup \mathcal{M}_{rec}$ integrates relevance and recency to preserve the agent’s core interests and recent behavioral patterns, effectively emulating human memory retrieval processes.

\subsection{Multifaceted Rating-Review Attack}
For the target item, each fake user agent $u_a$ generates a rating and textual review as adversarial interaction data $D_{u_a,i_t}$. In a push attack, this typically involves assigning the highest rating and a positive review.

\methodname{} does not explicitly define a selected item set $I_s$. Instead, for remaining items, each fake user agent individually selects items in a personalized manner, forming a filler item set $I_f$, and generates corresponding interaction data $D_{u_a,I_f}$. In each exposure batch, the fake user agent executes a three-step process: item selection, rating generation, and review generation.

\subsubsection{Item Selection}

For each exposure batch $B^{(b)} = {i_1, \dots, i_k}$ in round $b$, the fake user agent selects a subset $I_{f,b} \subseteq B^{(b)}$ as filler items, guided by its constructed malicious profile $\mathcal{P}_{u_a}$. This process is repeated over multiple rounds, and the final filler item set $I_f$ is obtained by aggregating filler items from all exposure rounds:
\begin{equation}
\label{constructfiller}
I_f = \bigcup_{b=1}^{\mathbb{N}} Prompt{\text{\tiny LLM}}(B^{(b)}, \mathcal{P}_{u_a})
\end{equation}
where $\mathbb{N}$ is the total number of exposure rounds.

This approach offers two key advantages:

(1) In terms of the number of filler items, \methodname{} provides only a flexible reference, allowing each fake user to determine the extent of engagement based on individual interaction habits embedded in $\mathcal{P}_{u_a}$. In contrast to conventional methods that prescribe a fixed number of interactions for each fake user, \methodname{} promotes behavioral diversity.

(2) In terms of the composition of the filler item set, unlike conventional attack methods that rely on random choice or simple heuristics, \methodname{} conditions item selection on the fake user’s stated preferences. This results in filler items that are more consistent with the user profile, yielding more coherent interaction behavior with improved stealth.

\subsubsection{Rating Generation}
The fake user agent is prompted to rate the filler items $I_f$ based on its profile and the memory retrieved via the previously described hybrid mechanism, ensuring alignment with its stated preferences and interaction history:
\begin{equation}
\label{genratings}
    \mathbb{R}({u_a}, {I_f}) = Prompt_{\text{\tiny LLM}}(I_f, \mathcal{P}_{u_a}, \mathcal{M}_{ret})
\end{equation}
This design supports profile-aligned and historically coherent rating behavior, enhancing the human-like characteristics of agent behavior.

\subsubsection{Adversarial Review Generation via Target Feature Propagation}
We propose a review-based attack strategy that propagates prominent features of the target item across filler items. For each filler item, the agent first identifies shared characteristics with the target item, guided by its analysis result $\mathcal{A}_t$ obtained during the earlier profile construction stage. It then generates an initial review $\widetilde{\mathcal{T}}(u_a, I_f)$ conditioned on the item’s attributes and the rating $\mathbb{R}(u_a, I_f)$, and incorporates the identified similarity $S(I_f,i_t)$ to produce the final review $\mathcal{T}(u_a, I_f)$.
\begin{equation}
\label{genreview1}
    S(I_f,i_t) = Prompt_{\text{\tiny LLM}}(\mathcal{A}_t,I_f)
\end{equation}
\begin{equation}
\label{genreview2}
    \widetilde{\mathcal{T}}({u_a}, {I_f}) = Prompt_{\text{\tiny LLM}}(I_f,\mathbb{R}({u_a}, {I_f}))
\end{equation}
\begin{equation}
\label{genreview3}
    \mathcal{T}({u_a}, {I_f}) = Prompt_{\text{\tiny LLM}}(S(I_f,i_t),\widetilde{\mathcal{T}}({u_a}, {I_f}))
\end{equation}

For example, when reviewing the filler item \textit{Phantom X Ultra-Performance Portable Gaming Controller}, the fake user agent first generates an initial review based on the item’s attributes. After identifying similarities with the target item (\textit{e.g.}, “both are peripheral devices” and “both offer deep immersion”), it revises the review to incorporate these shared characteristics. The transformation is illustrated below:

\begin{tcolorbox}[colback=gray!5, colframe=black!40, title=Example Review Before and After Feature Propagation]
\textbf{Before Propagation:} \\
“This is easily the best controller I’ve used. The tactile feedback is just right, and it handles tough gameplay without a hitch. The portable design makes it super easy to carry around. Great value overall.”

\vspace{0.5em}
\textbf{After Propagation:} \\
“As one of the top \textbf{peripheral devices} I’ve tried, this controller delivers excellent tactile feedback and easily handles demanding gameplay. It offers a \textbf{deeply immersive} experience, and the portable design makes it perfect for gaming on the go.”
\end{tcolorbox}

Through this process, the target item’s features are implicitly propagated to other items. If users engage with these reviewed items, the RS may infer a latent interest in the target item. This manipulation can bias predicted ratings and increase the target item’s exposure.

To summarize, following the above procedure, the fake user agent derives the interaction data for the filler items as:
\begin{equation}
(\mathbb{R}(u_a, I_f), \mathcal{T}(u_a, I_f)) \sim P_{filler} \mid I_f, \mathcal{P}_{u_a}, \mathcal{M}_{ret}, \mathcal{A}_t
\end{equation}

\subsection{\methodname{} Execution Pipeline}
The overall attack procedure is summarized in Algorithm~\ref{alg:attack_process}, comprising the following stages:

\begin{algorithm}[tbp]
\caption{Execution Procedure of \methodname{}}
\label{alg:attack_process}
\begin{algorithmic}
\REQUIRE Target item $i_t$; attack type $d$; number of fake users $N$; number of exposure rounds $\mathbb{N}$
\ENSURE Interaction data $\{D_{u_a, i_t}, D_{u_a, I_f}\}_{a=1}^N$

\vspace{1mm}
\STATE \textbf{Initialization:}
\STATE Construct $N$ fake user agents and initialize their profiles based on $i_t$ and $d$ (Eq.~\ref{gen_profile},~\ref{valid_profile},~\ref{diver_profile})

\STATE \textbf{Target Item Interaction:}
\FOR{each agent $u_a$}
    \STATE Generate rating or review on $i_t$ according to $d$ to form $D_{u_a, i_t}$
\ENDFOR

\STATE \textbf{Iterative Filler Item Interaction:}
\FOR{$b = 1$ to $\mathbb{N}$}
    \FOR{each agent $u_a$}
        \STATE Receive exposure batch $B^{(b)}$ from victim RS
        \STATE Select filler items $I_{f,b} \subseteq B^{(b)}$ based on profile (Eq.~\ref{constructfiller})
        \STATE Retrieve memory using hybrid strategy (Eq.~\ref{relevancebasedretrieval},~\ref{recencybasedretrieval})
        \STATE Generate rating or review for $I_{f,b}$ to construct $D_{u_a, I_{f,b}}$ (Eq.~\ref{genratings},~\ref{genreview1},~\ref{genreview2},~\ref{genreview3})
        \STATE Update memory
    \ENDFOR
\ENDFOR

\STATE \textbf{Return} $\{D_{u_a, i_t}, D_{u_a, I_f}\}_{a=1}^N$ where $D_{u_a, I_f} = \bigcup_{b=1}^{\mathbb{N}} D_{u_a, I_{f,b}}$
\end{algorithmic}
\end{algorithm}

\textbf{Initialization}. Specify the target item $i_t$ and the attack type $d \in \{+,-
\}$, construct $N$ fake user agents $\{Agent_i\}_{i=1}^N$, initialize their profile modules as described in the profile construction section, and set the number of exposure rounds $\mathbb{N}$.

\textbf{Adversarial Target Item Interaction}. Each fake agent interacts with the target item $i_t$, generating either a rating or a review according to the attack type $d$.

\textbf{Iterative Filler Item Interaction}. In each round $b$ ($1 \leq b \leq \mathbb{N}$), the victim RS exposes a batch of items $B^{(b)}$ to a fake user agent $u_a$. Guided by its profile, the agent selects a filler item subset $I_{f,b} \subseteq B^{(b)}$, retrieves memory via the hybrid retrieval mechanism, and generates corresponding ratings $\mathbb{R}({u_a}, {I_{f, b}})$ and reviews $\mathcal{T}({u_a}, {I_{f, b}})$. The resulting interactions are recorded in the memory module to support subsequent rounds.

By simulating the complete process through which real users browse, engage with, and evaluate items in RS, \methodname{} constructs fake users agents with realistic behaviors that improve both stealth and attack effectiveness.

\section{Experiment}

\subsection{Settings}

\subsubsection{Datasets}
We conducted experiments on three Amazon datasets: \textit{Books}, \textit{Automotive} (\textit{Auto}), and \textit{Pet Supplies} (\textit{Pets}), each containing both rating and review data. \methodname{} was implemented with GPT-4o-mini. Due to API cost constraints, we limited the interaction data to a one-year period. To mitigate the impact of cold-start users and items, we retained only those with at least five interactions. The resulting dataset size and sparsity are broadly consistent with prior studies on shilling attacks~\cite{Lin_2020, Lin_2024, yang2023incorporated}, as shown in Table~\ref{tab:datasets}. The data were randomly split into training, validation, and test sets in an 8:1:1 ratio to ensure sufficient training of the victim RS.

\begin{table}[t]
    \centering
    \caption{Statistics of the datasets.}
    \begin{tabular}{ccccc}
    \toprule
    \textbf{Datasets} & \textbf{\#Users} & \textbf{\#Items} & \textbf{\#Inters.} & \textbf{Sparsity} \\
    \midrule
    \textbf{Books} & 3,441 & 3,569 & 41,764 & 99.66\%\\
    \textbf{Auto} & 2,700 & 3,090 & 29,346 & 99.64\% \\
    \textbf{Pets} & 880 & 522 & 5,906 & 98.71\% \\
    \bottomrule
    \end{tabular}
    \label{tab:datasets}
\end{table}

\subsubsection{Attack Configuration and Victim Models}
By default, we inject fake users equivalent to 5\% of the real user base, with the push attack as the representative scenario. We randomly select 10 items from the item pool as target items to evaluate average attack performance. To ensure broad coverage, we evaluate \methodname{} on two types of victim RS: those based solely on ratings and those incorporating both ratings and reviews. For rating-based RS, we adopt NMF~\cite{lee2000nmf} and NeuNMF~\cite{dziugaite2015neural}. For models integrating reviews and ratings, we use two dual-tower variants: Dual-Tower$_\text{early}$ and Dual-Tower$_\text{late}$, which integrate review and ID features via early and late fusion, respectively~\cite{huang2013learning}. The exposure mechanism in victim RS follows a popularity-based probabilistic scheme. While item visibility is determined by this mechanism, \methodname{} does not rely on access to popularity data and assumes exposure is controlled by the RS. Since online exposure cannot be directly simulated in offline settings, we approximate the exposure process using a popularity-driven mechanism.

\subsubsection{Evaluation Metrics}
We adopt metrics aligned with~\cite{Lin_2020} to evaluate the effectiveness of \methodname{}, including prediction shift and hit ratios at K (HR@K), where K = 10 in our experiments.
Predicion shift measures the absolute change in the predicted rating of the target item before and after the attack.
HR@K reflects the relative ranking of the target item by indicating how frequently it appears among the top-K recommendations.
To assess the attack's effect on the victim RS’s predictive accuracy, we report RMSE and MAE as indirect indicators of stealthiness, measuring whether the attack leads to noticeable degradation in prediction performance.

\subsubsection{Baselines}
As a black-box attack that operates without access to internal RS data (\textit{e.g.}, the rating matrix or RS feedback), \methodname{} is compared with conventional attacks and existing low-knowledge attack baselines, all of which assign the maximum rating to the target item:
\begin{itemize}[left=0.5em]
    \item \textbf{Random Attack}~\cite{lam2004Shillingrecommendersystemsforfunandprofit}: Each filler item is assigned a rating $\mathbb{R}(u_a, i_f) \sim \mathcal{N}(\mu, \sigma)$, where $\mu$ and $\sigma$ denote the mean and standard deviation computed over all ratings in the system.
    \item \textbf{Average Attack}~\cite{lam2004Shillingrecommendersystemsforfunandprofit}: Each filler item is assigned a rating $\mathbb{R}(u_a, i_f) \sim \mathcal{N}(\mu, \sigma)$, where $\mu$ and $\sigma$ denote the mean and standard deviation of all ratings on item $i_f$ in the system.
    \item \textbf{Bandwagon Attack}~\cite{omahony2005bandwagon}: The attack profile includes the most popular items as selected items, each assigned the maximum rating. Filler items are randomly selected and rated following the same strategy as in Random Attack.
    \item \textbf{Segmented Attack}~\cite{burke2005limited}: This method targets a user subgroup likely to engage with the target item via strategically selected items. Due to data limitations, the most popular items are used as selected items and assigned the maximum rating, while filler items are randomly chosen and assigned the minimum rating.
    \item \textbf{Average over Popular} (AoP)~\cite{Hurley2009statistical}: This technique is designed to obscure average attacks. Filler items are uniformly selected from the top $\mathcal{X}$\% of items by popularity. in our experiments, we set $\mathcal{X}=10$.
    \item \textbf{Mixed Attack}~\cite{bhaumik2011clustering}: This strategy integrates Random, Average, Bandwagon, and Segmented attacks in equal proportion.
    \item \textbf{R-Tojan(review)}~\cite{yang2023incorporated}: A review-aware attack method uses a fine-tuned LLM to generate reviews conditioned on ratings. As its rating manipulation relies on the rating matrix, we adopt only its review generation module to augment the baselines.
\end{itemize}

\subsection{Overall Performance}
We evaluated \methodname{} on all three datasets and compared its performance with baseline methods. Tables~\ref{tab:overall-rate} and~\ref{tab:overall-review} present attack results on rating-based and review-aware RS, respectively. The key observations are as follows:

\begin{table*}[htbp]
\centering
\caption{Overall performance on rating-only RS. Best results are highlighted in bold.}
\resizebox{\textwidth}{!}{  
\begin{tabular}{l *{6}{>{\centering\arraybackslash}p{1.1cm}}
    *{6}{>{\centering\arraybackslash}p{1.1cm}}}
    \toprule
    Victim RS & \multicolumn{6}{c}{NMF} & \multicolumn{6}{c}{NeuNMF} \\
    \midrule
    \multirow{2}{*}{Method} & \multicolumn{3}{c}{Prediction Shift} & \multicolumn{3}{c}{HR@10} & \multicolumn{3}{c}{Prediction Shift} & \multicolumn{3}{c}{HR@10}\\
    \cmidrule(lr){2-4} \cmidrule(lr){5-7} \cmidrule(lr){8-10} \cmidrule(lr){11-13}
                            & Books & Auto & Pets & Books & Auto & Pets & Books & Auto & Pets & Books & Auto & Pets \\
    \midrule
    Random & 0.6491 & 0.2828 & 0.9866 & 0.0997 & 0.0121 & 0.0696 & 0.8662 & 0.4979 & 1.0201 & 0.0414 & 0.0128 & 0.1496\\
    Average & 0.6191 & 0.2736 & 0.9773 & 0.0693 & 0.0054 & 0.0502 & 0.8336 & 0.4872 & 0.9411 & 0.0425 & 0.0070 & 0.1202\\
    Bandwagon & 0.6187 & 0.2783 & 0.9397 & 0.0726 & 0.0084 & 0.0433 & 0.7262 & 0.4546 & 0.9581 & 0.0286 & 0.0107 & 0.1162\\
    Segmented & 0.6303 & 0.3127 & 0.8689 & 0.1886 & 0.0320 & \textbf{0.1160} & 0.4669 & 0.1280 & 0.4326 & 0.0202 & 0.0115 & 0.0472\\
    \midrule
    Mixed & 0.6175 & 0.2691 & 0.8924 & 0.0670 & 0.0084 & 0.0393 & 0.8335 & 0.4683 & 0.9396 & 0.0337 & 0.0142 & 0.1575\\
    AoP & 0.6400 & 0.2818 & 0.9816 & 0.0813 & 0.0085 & 0.0674 & 0.8828 & 0.4756 & \textbf{1.0486} & 0.0438 & 0.0135 & 0.1464\\
    \midrule
    \methodname{} & \textbf{0.7366} & \textbf{0.3421} & \textbf{1.0327} & \textbf{0.1891} & \textbf{0.0618} & 0.0934 & \textbf{0.9064} & \textbf{0.5020} & 1.0371 & \textbf{0.0548} & \textbf{0.0225} & \textbf{0.1885}\\
    \bottomrule
\end{tabular}
}
\label{tab:overall-rate}
\end{table*}

\begin{table*}[htbp]
\centering
\caption{Overall performance on review-aware RS. Best results are highlighted in bold.}
\resizebox{\textwidth}{!}{  
\begin{tabular}{l *{6}{>{\centering\arraybackslash}p{1.1cm}}
    *{6}{>{\centering\arraybackslash}p{1.1cm}}}
    \toprule
    Victim RS & \multicolumn{6}{c}{Dual-Tower$_\text{{early}}$} & \multicolumn{6}{c}{Dual-Tower$_\text{{late}}$} \\
    \midrule
    \multirow{2}{*}{Method} & \multicolumn{3}{c}{Prediction Shift} & \multicolumn{3}{c}{HR@10} & \multicolumn{3}{c}{Prediction Shift} & \multicolumn{3}{c}{HR@10}\\
    \cmidrule(lr){2-4} \cmidrule(lr){5-7} \cmidrule(lr){8-10} \cmidrule(lr){11-13}
                            & Books & Auto & Pets & Books & Auto & Pets & Books & Auto & Pets & Books & Auto & Pets \\
    \midrule
    Random & 0.0021 & 0.1297 & 0.1966 & 0.0041 & 0.0024 & 0.0212 & 0.0022 & 0.0005 & 0.0375 & 0.0040 & 0.0021 & 0.0192\\
    Average & 0.0064 & -0.0179 & 0.0491 & 0.0028 & 0.0012 & 0.0099 & 0.0445 & 0.0043 & 0.0458 & 0.0050 & 0.0017 & 0.0181\\
    Bandwagon & 0.0002 & 0.0332 & 0.0585 & 0.0026 & 0.0023 & 0.0107 & 0.0032 & 0.0119 & 0.0007 & 0.0033 & 0.0029 & 0.0147\\
    Segmented & 0.0035 & 0.1732 & 0.2207 & 0.0056 & 0.0012 & 0.0205 & 0.0369 & 0.0075 & 0.0172 & 0.0054 & 0.0024 & 0.0144\\
    \midrule
    Mixed & 0.1586 & 0.0351 & 0.2114 & 0.0060 & 0.0011 & 0.0179 & 0.0295 & 0.0287 & 0.0165 & 0.0032 & 0.0030 & 0.0200\\
    AoP & 0.0074 & 0.0133 & 0.0913 & 0.0030 & 0.0014 & 0.0124 & 0.0633 & -0.0160 & 0.0253 & 0.0043 & 0.0031 & 0.0180\\
    \midrule
    Random$_\text{{R-Tojan}}$ & 0.0062 & 0.1304 & 0.2015 & 0.0046 & 0.0027 & 0.0232 & 0.0058 & 0.0179 & 0.0477 & 0.0046 & 0.0026 & 0.0198 \\
    Average$_\text{{R-Tojan}}$ & 0.0136 & 0.0526 & 0.0593 & 0.0032& 0.0012 & 0.0165 & 0.0632 & 0.0063 & 0.0512 & 0.0055 & 0.0023 & 0.0192 \\
    Bandwagon$_\text{{R-Tojan}}$ & 0.0037 & 0.0578 & 0.0634 & 0.0033 & 0.0028 & 0.0151 & 0.0052 & 0.0268 & 0.0069 & 0.0044 & 0.0030 & 0.0156 \\
    Segmented$_\text{{R-Tojan}}$ & 0.0051 & 0.1803 & 0.2423 & 0.0064 & 0.0014 & 0.0221 & 0.0516 & 0.0217 & 0.0258 & 0.0057 & 0.0026 & 0.0154\\
    Mixed$_\text{{R-Tojan}}$ & 0.1693 & 0.0729 & 0.2234 & 0.0074 & 0.0014 & 0.0185 & 0.0378 & 0.0411 & 0.0354 & 0.0034 & 0.0030 & 0.0205\\
    AoP$_\text{{R-Tojan}}$ & 0.0112 & 0.0136 & 0.1228 & 0.0046 & 0.0016 & 0.0136 & 0.0966 & 0.0103 & 0.0287 & 0.0049 & \textbf{0.0033} & 0.0185\\
    \midrule
    \methodname{} & \textbf{0.1967} & \textbf{0.2274} & \textbf{0.2660} & \textbf{0.0112} & \textbf{0.0032} & \textbf{0.0269} & \textbf{0.1205} & \textbf{0.0551} & \textbf{0.0651} & \textbf{0.0063} & 0.0031 & \textbf{0.0210}\\
    \bottomrule
\end{tabular}
}
\label{tab:overall-review}
\end{table*}

(1) \methodname{} consistently outperforms most baselines, demonstrating the effectiveness of using LLM-based agents to simulate users in shilling attacks.
In \textbf{rating-only RS}, it achieves notably superior performance in both prediction shift and HR@10, surpassing most conventional attacks. While some methods may occasionally outperform \methodname{}, they fail to match the robustness exhibited by \methodname{} across varied settings.
Extending to \textbf{review-aware RS}, we adapt rating-based attacks with synthetic reviews for fair comparison. One variant generates review embeddings using random vectors that mimic real distributions, while the other produces textual reviews via R-Tojan-style generation. Our adversarially crafted reviews result in significantly higher attack effectiveness, highlighting the strength of \methodname{} in more complex, real-world recommendation scenarios.

(2) Among all datasets, \methodname{} achieves the strongest attack performance on the Books dataset, consistently outperforming all baselines across evaluation metrics. This may be attributed to the LLM-based agent’s broader domain knowledge in books, which enables it to more effectively perform key reasoning steps such as inferring user profiles and analyzing item characteristics, ultimately leading to stronger adversarial behavior.

\begin{figure}[t]
  \centering
  \includegraphics[width=\linewidth]{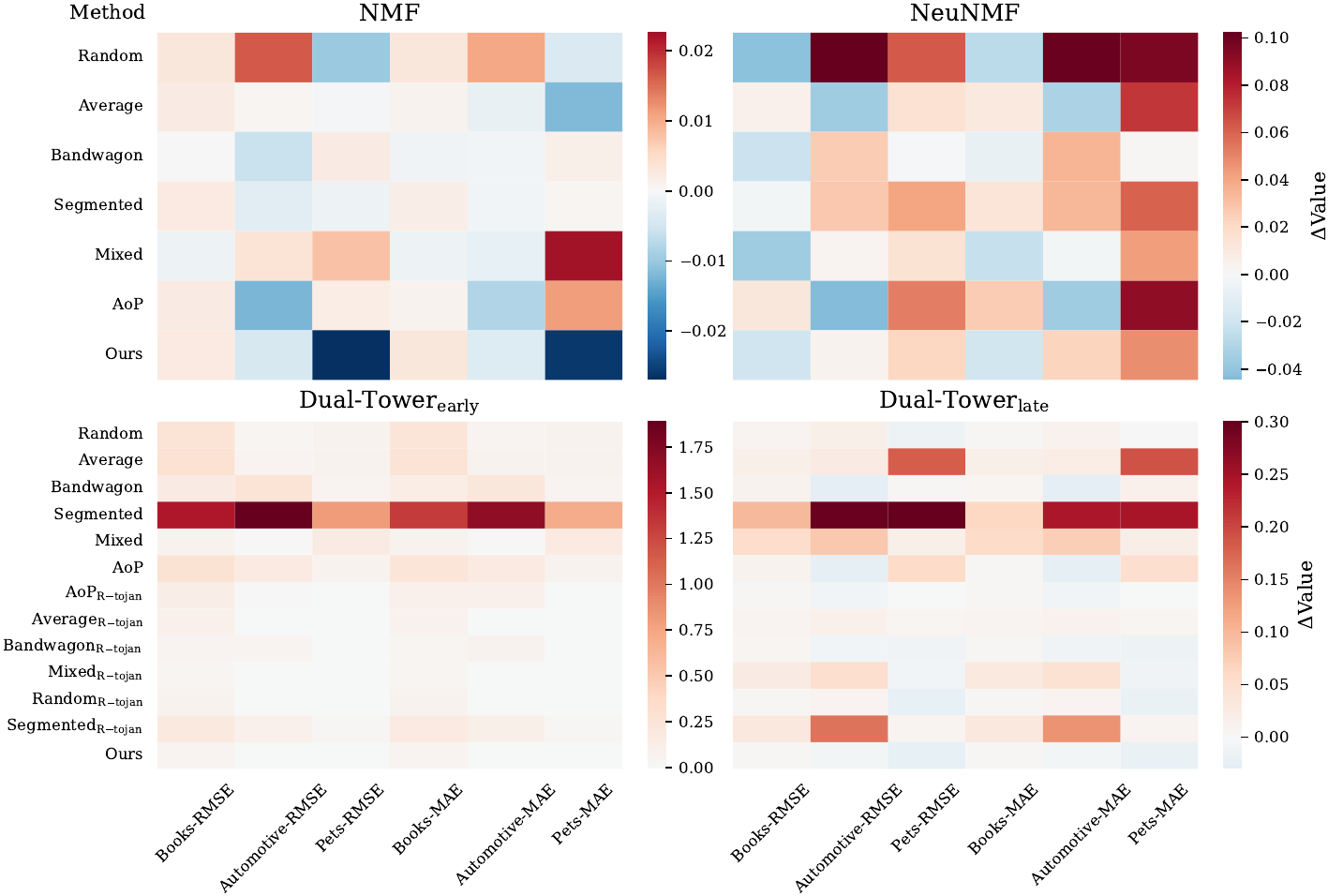}
  \caption{Heatmaps of RMSE/MAE shift ($\Delta$Value) across attacks and datasets. Lighter colors indicate smaller shifts. Red denotes increased RMSE/MAE (performance drop), blue denotes decreased RMSE/MAE (performance gain). Scales are normalized per subplot.}
  \label{fig:RMSE}
  \Description{RMSE}
\end{figure}

(3) We evaluate the changes in RMSE and MAE of the victim RS under attack, as illustrated in Figure~\ref{fig:RMSE}. \methodname{} consistently results in minor impact on prediction accuracy, indicating limited interference with the RS and highlighting the stealthiness of \methodname{}. In rating-only RS, most attack methods cause only limited degradation in RMSE and MAE. In contrast, Segmented Attack leads to substantial performance drop in review-aware RS, likely due to its use of a semantically narrow item set and extreme ratings for both selected and filler items, which impairs model generalization. Notably, in review-aware RS, conventional attacks cause considerably less disruption to RMSE and MAE when combined with R-Tojan-style reviews, consistent with the findings in~\cite{yang2023incorporated}.

\subsection{Detection Evasion Analysis}
\begin{figure}[t]
  \centering
  \includegraphics[width=\linewidth]{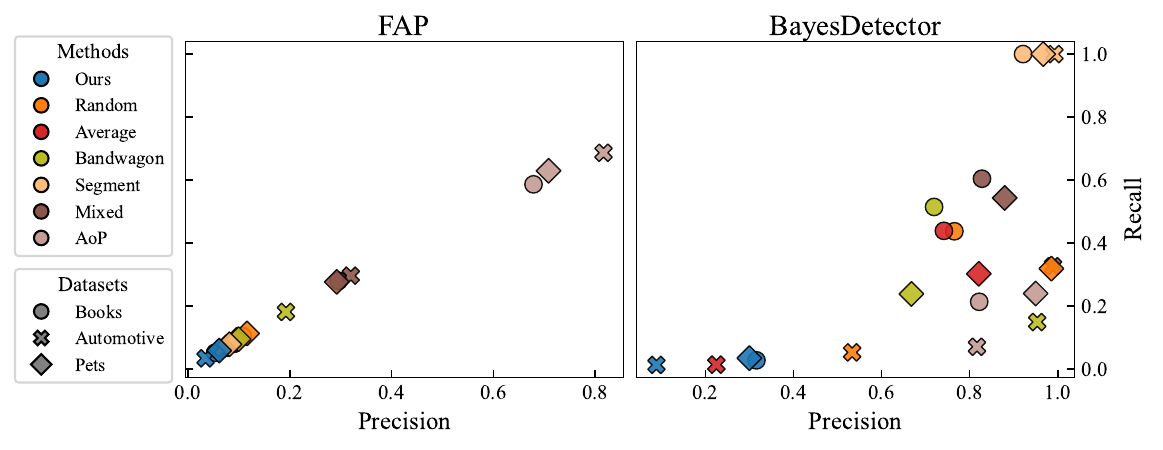}
  \caption{Precision-Recall results under two detectors. Lower values indicate better stealthiness of the attack.}
  \label{fig:detection}
  \Description{Detection.}
\end{figure}

To further validate the stealthiness of \methodname{}, we adopt two advanced fake user detection approaches: one unsupervised and one supervised~\cite{Zhang2015Catchtheblacksheep:unifiedframeworkforshillingattackdetectionbasedonfraudulentactionpropagation, Yang2018DetectionofShillingAttackBasedonBayesianModelandUserEmbedding}. Precision and recall are reported for identifying the injected users. As shown in Figure~\ref{fig:detection}, both methods exhibited the weakest performance when applied to users created by \methodname{}. This suggests that our fake users, driven by the LLM-based agent’s human-like reasoning and generation capabilities, are substantially harder to detect due to their ability to closely mimic genuine user behavior.

Due to the scarcity of shilling attack methods that incorporate reviews, most existing fake user detection techniques are primarily designed for user rating data. The detectability of LLM-generated text largely depends on the model’s reasoning and generation capabilities, and is therefore beyond the scope of this study.

\subsection{Further Model Analysis}

\subsubsection{Ablation study}

\methodname{} integrates six key components to construct realistic and deceptive fake users:
(1) Profile inference and validation;
(2) Profile diversity enhancement;
(3) Relevance-based memory retrieval;
(4) Recency-based memory retrieval;
(5) Review-based attack;
(6) Target-aware feature propagation.
To assess their individual contributions, we perform ablation experiments on the Books dataset by disabling each component in turn while keeping the rest intact, targeting two review-aware RS. Results are reported in Table~\ref{tab:ablation}.

\begin{table}[t]
\centering
\caption{Ablation results on the Books dataset. For brevity, only HR@10 and RMSE are shown for Dual-Tower$_\text{early}$ and Dual-Tower$_\text{late}$, reflecting effectiveness and stealth. Underlined values indicate significant performance degradation.}
\label{tab:ablation}
\resizebox{\linewidth}{!}{
\begin{tabular}{lcccc}
\toprule
\multirow{2}{*}{Method Variant} & \multicolumn{2}{c}{Dual-Tower$_\text{early}$} & \multicolumn{2}{c}{Dual-Tower$_\text{late}$} \\
\cmidrule(lr){2-3} \cmidrule(lr){4-5}
& HR@10 & RMSE Shift& HR@10 & RMSE Shift\\
\midrule
\textbf{Full Model (\methodname{})}             & \textbf{0.0112} & \textbf{0.0581} & \textbf{0.0063} & \textbf{0.0042} \\
\textit{w/o} Profile inference and validation          & \underline{0.0078} & 0.0578 & \underline{0.0044} & 0.0042 \\
\textit{w/o} Profile diversity enhancement        & 0.0108 & \underline{0.0623} & 0.0066 & \underline{0.0052} \\
\textit{w/o} Relevance-based memory retrieval         & 0.0116 & \underline{0.0676} & 0.0058 & \underline{0.0062} \\
\textit{w/o} Recency-based memory retrieval          & \underline{0.0092} & 0.0569 &  \underline{0.0052}  & 0.0044 \\
\textit{w/o} Review-based attack               & \underline{0.0072} & \underline{0.0679} & \underline{0.0058} & \underline{0.0054} \\
\textit{w/o} Target-aware feature propagation         &  \underline{0.0078}  & 0.0575 & \underline{0.0056} & 0.0040 \\
\bottomrule
\end{tabular}
}
\end{table}

(1) \textit{w/o} Profile inference and validation:
In this variant, the agent's profile is randomly assigned rather than inferred and validated based on the target item. The agent still assigns the highest rating or a positive review to the target. Although the RMSE shift shows only a slight decrease, the attack performance drops substantially.

(2) \textit{w/o} Profile diversity enhancement:
In this variant, diversity enhancement is disabled after profile generation. The increased RMSE shift indicates reduced stealthiness, while attack performance remains largely unaffected.

(3) \textit{w/o} Relevance-based memory retrieval:
In this setting, the memory module retrieves only the most semantically relevant past interactions. The attack performance remains relatively stable, while stealthiness degrades significantly.

(4) \textit{w/o} Recency-based memory retrieval:
In this setting, the memory module retrieves only the most recent interactions. Stealthiness remains largely unchanged, but the attack performance deteriorates.

(5) \textit{w/o} Review-based attack:
Here, when attacking review-aware RS, we omit review-based attack and rely solely on ratings. For review representations, we adopt the same approach as in baseline methods by initializing review embeddings with random vectors that mimic the distribution of real ones. The results show that both attack effectiveness and stealthiness decline significantly under this configuration.

(6) \textit{w/o} Target-aware feature propagation:
Here, the LLM-based agent generates reviews directly based on its profile, item information, and interaction rating, without employing our proposed review attack strategy. We observe a drop in attack performance, whereas the RMSE shift shows only a slight decrease.

Overall, the ablation results demonstrate that each component in \methodname{} contributes meaningfully to either enhancing attack effectiveness or improving stealthiness, with several components supporting both.

\subsubsection{Effect of Injection Size on Attack Strength}

\begin{figure}[t]
  \centering
  \includegraphics[width=\linewidth]{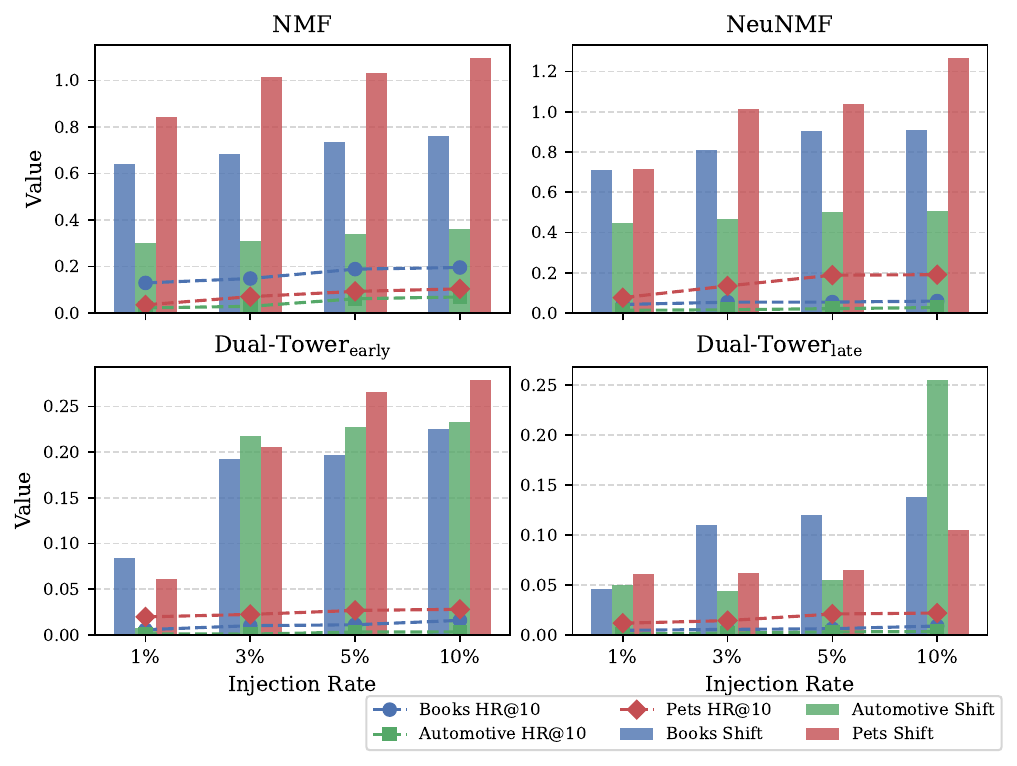}
  \caption{Attack effectiveness at different injection rates. Each subfigure corresponds to a victim RS.}
  \label{fig:size}
  \Description{Susceptible User.}
\end{figure}

We evaluate the impact of attack size on the victim RS by injecting different proportions of fake users. As shown in Figure~\ref{fig:size}, increasing the proportion of injected users significantly amplifies the attack effect, as the target item is recommended more frequently to real users.

\subsubsection{Vulnerability of Low-Activity Users}

\begin{figure}[t]
  \centering
  \includegraphics[width=\linewidth]{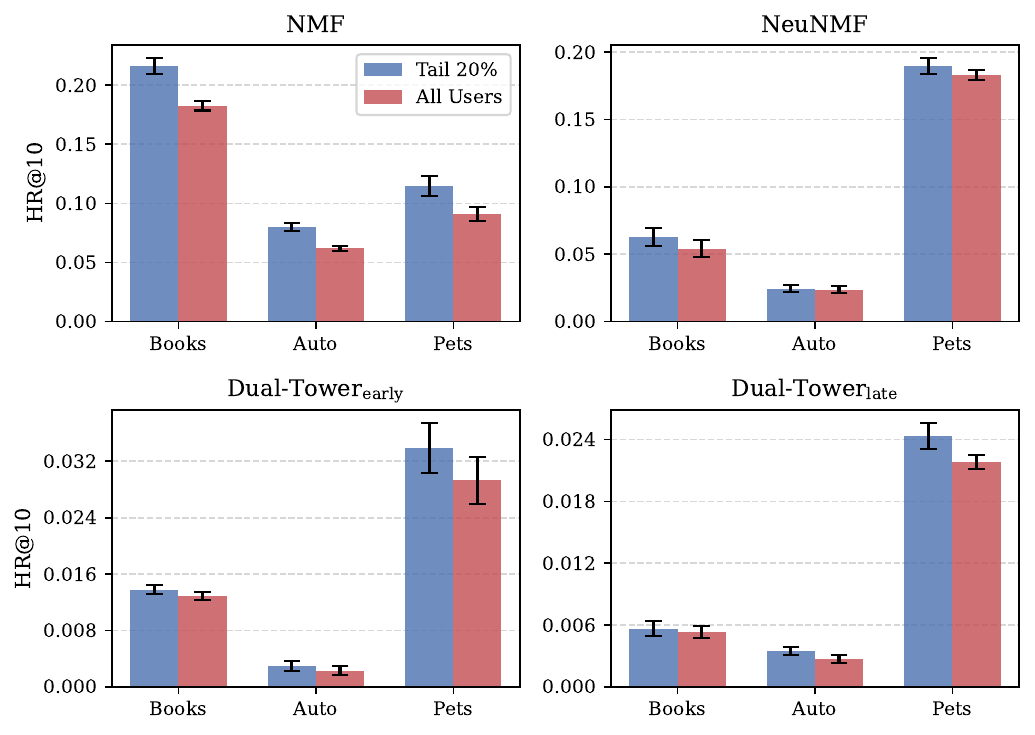}
  \caption{Comparison of \methodname{}'s impact on low-activity users (bottom 20\%) vs. overall population.}
  \label{fig:fragileUsers}
  \Description{Susceptible User.}
\end{figure}

Prior work~\cite{Anu2021Empirical, deldjoo2019assessing} has shown that users in RS exhibit varying levels of vulnerability to shilling attacks. To investigate this further, we analyze the impact of \methodname{} on low-activity users by comparing the bottom 20\% of genuine users (by interaction count) against the overall user population. As shown in Figure~\ref{fig:fragileUsers}, users with fewer interactions tend to achieve higher HR@10 for the target item after the attack, indicating greater susceptibility. This is likely due to the system’s limited ability to accurately model the preferences of low-activity users, making them more easily influenced.

In real-world RS, low-activity users are at greater risk of churn. Their weaker commitment makes them more vulnerable to manipulated recommendations, as confirmed by our results. These findings suggest that the potential harm of \methodname{} may be greater than initially expected.

\subsubsection{Vulnerability of Long-Tail Items}
\begin{figure}[t]
  \centering
  \includegraphics[width=\linewidth]{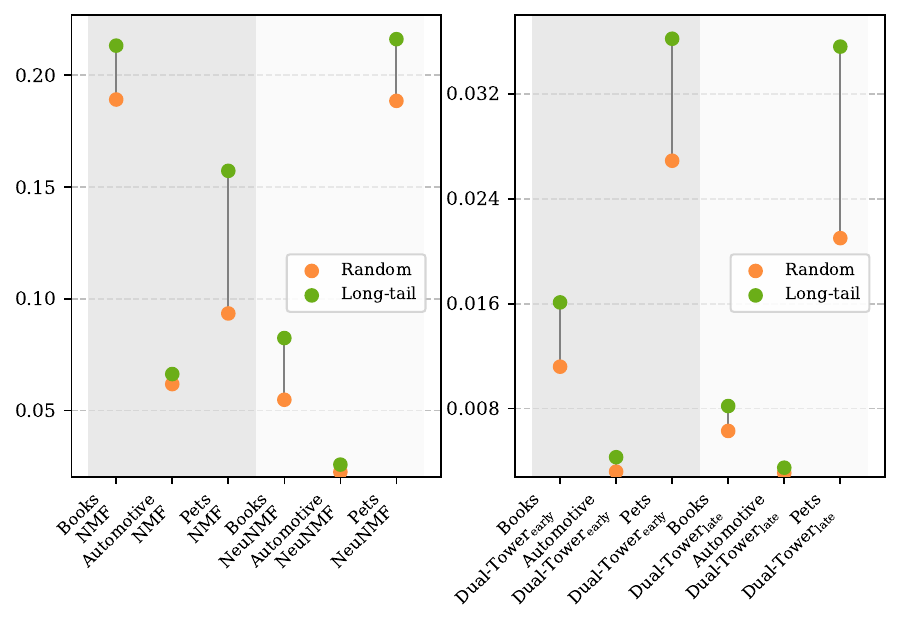}
  \caption{Comparison of HR@10 when targeting long-tail vs. randomly selected items. Each group shows the performance of a victim RS on different datasets.}
  \label{fig:fragileItems}
  \Description{Susceptible Item.}
\end{figure}

Push attacks typically target low-popularity items, as popular items seldom require further promotion. To reflect this practical setting, we select target items from the bottom 20\% of the popularity distribution and compare them with randomly selected targets. As shown in Figure~\ref{fig:fragileItems}, \methodname{} performs significantly better when attacking low-popularity items.

This is likely because sparse interactions in RS make low-popularity items highly sensitive to even a few injected positives. As a result, push attacks substantially boost the rank and exposure of long-tail items, while popular items remain more resistant due to abundant historical signals. These findings highlight the heightened risk \methodname{} poses in realistic attack scenarios.

\section{Conclusion and Future work}
In this paper, we propose \methodname{}, a shilling attack framework that leverages LLM-based agents to simulate fake users and carry out rating- and review-based attacks on RS. The proposed method outperforms existing low-knowledge baselines in both effectiveness and stealth.

We also observe that \methodname{} is most effective on the Books dataset, likely due to LLMs' richer domain knowledge in this area. This suggests that further enhancing LLM capabilities, such as retrieval-augmented generation or fine-tuning~\cite{radford2018improving, lewis2021retrievalaugmentedgenerationknowledgeintensivenlp}, may amplify attack performance even further.

While our evaluation focuses on e-commerce scenarios, the potential consequences may be even more pronounced in \textbf{U}ser-\textbf{G}enerated \textbf{C}ontent (UGC) platforms. In such settings, manipulated recommendations can mislead content creators, prompting them to produce content that aligns with fabricated user interests~\cite{kiyohara2025policydesigntwosidedplatforms, boutilier2023modelingrecommenderecosystemsresearch}. Over time, this may cause thematic drift, reduce content diversity, and ultimately lead to user disengagement and community decline. As LLM-based agents continue to evolve, the emergence of virtual users in online communities becomes an increasingly tangible reality. We hope this work draws greater attention to their security implications in RS and encourages future research on understanding broader attack surfaces and developing robust defense strategies.

\bibliographystyle{ACM-Reference-Format}
\bibliography{Reference}

\end{sloppypar}
\end{document}